\documentclass[doublecol,figures]{epl2}

\title{Thermal rounding of the depinning transition
}

\author{S. Bustingorry\inst{1} \and A. B. Kolton\inst{2} \and T. Giamarchi\inst{1}}
\shortauthor{S. Bustingorry \etal}

\institute{
  \inst{1} DPMC-MaNEP, Universit{\'e} de Gen{\`e}ve, 24 Quai Ernest Ansermet, 1211 Gen{\`e}ve 4, Switzerland\\
  \inst{2} Dep. de F{\'i}sica At\'omica, Molecular y Nuclear,  Universidad Complutense de Madrid, 28040 Madrid, Spain
}
\pacs{64.60.Ht}{Dynamic critical phenomena}
\pacs{75.60.Ch}{Domain walls and domain structure}
\pacs{05.70.Ln}{Nonequilibrium and irreversible thermodynamics}

\abstract{
We study thermal effects at the depinning transition by numerical simulations of driven one-dimensional elastic interfaces in a disordered medium. We find that the velocity of the interface, evaluated at the critical depinning force, can be correctly described with the power law $v\sim T^\psi$, where $\psi$ is the thermal exponent. Using the sample-dependent value of the critical force, we precisely evaluate the value of $\psi$ directly from the temperature dependence of the velocity, obtaining the value $\psi = 0.15 \pm 0.01$. By measuring the structure factor of the interface we show that both the thermally-rounded and the $T=0$ depinning, display the same large-scale geometry, described by an identical divergence of a characteristic length with the velocity $\xi \propto v^{-\nu/\beta}$, where $\nu$ and $\beta$ are respectively the $T=0$ correlation and depinning exponents. We discuss the comparison of our results with previous estimates of the thermal exponent and the direct consequences for recent experiments on magnetic domain wall motion in ferromagnetic thin films.}

\begin{document}

\maketitle

Understanding the physics of disordered elastic lines has a direct impact on a large
variety of experimental systems in condensed matter. Indeed such systems are realized
for isolated lines by magnetic~\cite{lemerle_domainwall_creep,bauer_deroughening_magnetic2,yamanouchi_creep_ferromagnetic_semiconductor2,metaxas_depinning_thermal_rounding} or ferroelectric~\cite{paruch_ferro_roughness_dipolar,paruch_ferro_quench} domain walls, contact lines~\cite{moulinet_distribution_width_contact_line2} and fractures~\cite{bouchaud_crack_propagation,alava_review_cracks} and for periodic systems
by vortex lattices~\cite{blatter_vortex_review,du_aging_bragg_glass}, charge density waves~\cite{nattermann_cdw_review} or wigner crystals~\cite{giamarchi_electronic_crystals_review}.
In these systems, one particularly important question is to understand the response of the interface
to an externally applied force, such response being directly measurable in all the above systems.

In all these systems the competition between the disorder and the elasticity of the lines leads to
pinned configurations. In order to set the system in motion it is thus necessary to apply a force $F$ exceeding
a critical force $F_c$. At zero temperature the average velocity $V$ remains zero for $F<F_c$, while the system
moves for $F>F_c$. In addition to their experimental relevance, the understanding of the above properties
constitutes a considerable theoretical challenge \cite{giamarchi_domainwall_review}. A very fruitful approach was to draw on the analogies between such a phenomenon and a standard critical phenomenon to predict scaling properties of the various physical observables close to the depinning transition~\cite{fisher_depinning_meanfield}. In that respect a very important question is how to extend the above results to the case of finite temperature. This is of course directly relevant for the experimental systems. A finite temperature allows the system to move even if $F<F_c$ leading to a thermal rounding of the depinning transition.
Indeed, such rounding was recently experimentally observed~\cite{metaxas_depinning_thermal_rounding}. A natural way to analyze such thermal effect is to draw further on the analogy with critical phenomena. This is however not so simple since the depinning is after all an out of equilibrium feature and it is unclear how far the analogy with standard critical phenomena should be carried out. Recently, serious differences between depinning and a standard critical phenomenon were indeed pointed out~\cite{kolton_depinning_zerot2}.

It is thus crucial to perform a detailed analysis of the thermal rounding of
the depinning transition. Unfortunately, the analytical methods able to tackle
such an out of equilibrium situation are rare. A very successful method to
describe the zero temperature depinning is the functional renormalization
group
(FRG)~\cite{narayan_fisher_cdw,nattermann_stepanow_depinning,chauve_creep}. It
allowed to obtain an expansion in $4-d$, $d$ being the dimension of the
interface, of the depinning exponents for $T=0$. Recently, FRG equations
describing the full dynamics of the interface at finite temperature have been
derived~\cite{chauve_creep}. Such equations were proven to be very efficient
to obtain the dynamics at very small force (the so called creep behavior
\cite{blatter_vortex_review}). Unfortunately, solving them around depinning at
finite temperature is considerably more complicated, and no complete solution
exists to date, even if some crude prediction for the rounding exponent can be
made~\cite{tang_stepanow_thermal_rounding,chauve_creep}. In addition to the
intrinsic problem of solving the equations, the fact that they are obtained in
a $4-d$ expansion makes them poorly suited to quantitatively describe the
important case of the one dimensional interface. Alternative methods, such as
numerical studies are thus clearly needed. Indeed, there were various attempts
to obtain the thermal exponent from numerical simulations. In
ref.~\cite{chen_marchetti} the authors used a dynamically generated disorder
environment, corresponding to an infinite size system in the dimension where
the interface is allowed to fluctuate. On the other hand, the random field
Ising model at finite temperature was used to compute the thermal exponent in
various
dimensions~\cite{nowak_thermal_rounding,roters_thermal_rounding1,roters_thermal_rounding2}.
However in these works, in addition to the intrinsic numerical limitations,
the authors were forced to use the average critical values for the pinning
force, obtained by fitting the zero-temperature velocity-force
characteristics. Unfortunately, this force has severe sample-to-sample
fluctuations, which indeed strongly depend on the size of the
system~\cite{rosso_depinning_simulation,bolech_critical_force_distribution}.
This made the determination of the thermal rounding exponent obtained by such
methods relatively imprecise, and more importantly made it difficult to check
whether one had the proper scaling to even define such an exponent. In view of
the existing differences between depinning and a standard critical phenomena
it is indeed important to check whether such scaling exists.

In this paper we analyze the finite-temperature dynamics of driven elastic interfaces evolving in a two-dimensional random media. We use a conceptually different approach, where we use the pinning force for each disorder realization as it was done in~\cite{duemmer} for analyzing the zero-temperature case. Such force can be now determined with great accuracy~\cite{rosso_depinning_simulation}. We obtain a thermal exponent $\psi = 0.15 \pm 0.01$. Although we obtain the thermal exponent directly from the velocity versus temperature relation at the critical force, we show that this value can also account for finite size effects by assuming that the large-scale behavior is controlled by a temperature-dependent correlation length, growing with decreasing velocity in the same way as for the zero temperature case. The existence of such velocity-dependent correlation-length is directly proved by the analysis of the finite-temperature structure factor. This shows that the large-scale geometry of the interface is only controlled by the velocity regardless on whether one approaches the critical point from positive forces at zero temperature or from the temperature axis at the critical force.

We consider an elastic interface described by a single valued function $u(z,t)$, giving its transverse position $u$ in the $z$ axis. The interface evolves with time $t$ obeying the equation
\begin{equation}
\label{eq:EW-F}
\gamma \, \partial_t u(z,t) = c \, \partial^2_z u(z,t) + F_p(u,z) + F + \eta(z,t),
\end{equation}
where $\gamma$ is the friction coefficient and $c$ the elastic constant. The pinning force $F_p(u,z) = - \partial_u U(u,z)$ represents the effects of a random-bond disorder described by the potential $U(u,z)$, whose sample to sample fluctuations are given by $\overline{\left[ U(u,z)- U(u',z') \right]^2} = \delta(z-z') \, R(u-u')$, where $R(u)$ stands for a correlator of range $r_f$~\cite{chauve_creep}, and the overline indicates average over disorder realizations.
The thermal noise $\eta(z,t)$ satisfies $\langle \eta(z,t) \rangle =0$ and $\langle \eta(z,t) \eta(z',t') \rangle = 2 \gamma T \delta(t-t') \delta(u-u')$. Finally, the force $F$ corresponds to an uniform and constant external field.

In order to numerically solve eq.~(\ref{eq:EW-F}) we discretize the $z$
direction in $L$ segments of size $\delta z=1$, {\it i.e.} $z \to
j=0,...,L-1$, while keeping $u_j(t)$ as a continuous variable. Then the
equation is integrated using the Euler method with a time step $\delta
t=0.01$. The continuous random potential is modeled by a cubic spline passing
through $M$ regularly spaced uncorrelated Gaussian numbers
points~\cite{rosso_depinning_simulation,kolton_creep2}. The numerical
simulations are performed using $\gamma=1$, $c=1$, $r_f=1$, and with $R(0)=1$
giving the strength of the disorder. Periodic boundary conditions are used in
both spatial directions, thus defining a $L\times M$ system.

\begin{figure}
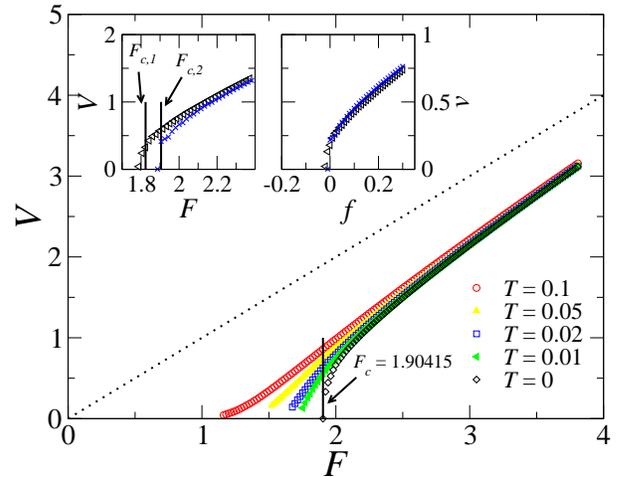

\onefigure[width=8cm]{VdeF.eps} \caption{Velocity $V$ versus force $F$ for a
given disorder realization and different temperatures. The system size is
$L=1024$, and the critical force $F_c$ is also indicated. The left-inset shows
the velocity-force curves for two different disorder realizations, with
critical forces $F_{c,1}=1.82395$ and $F_{c,2}=1.90555$ respectively, at a
fixed small temperature $T=0.001$ and for a small size $L=128$. The
right-inset shows the same data as in the left-inset but in reduced variables
$v=V/F_c$ and $f=(F-F_c)/F_c$. } \label{f:VdeF}
\end{figure}

Figure~\ref{f:VdeF} shows a typical velocity-force characteristics for a single sample of size $L=1024$, obtained by computing the steady-state velocity $V=\langle \partial_t u(z,t) \rangle$ through numerical integration of Eq.~(\ref{eq:EW-F}).
The sample-dependent critical force is also quoted, which can be obtained for each realization of the disorder configuration using a fastly-convergent algorithm~\cite{rosso_depinning_simulation}. We can observe that the sharp depinning transition is rounded by temperature as expected. The straight dashed-line indicates the fast-flow limit $V=F$, which is reached at high force. The left-inset shows the velocity-force curve for $T=0.001$, for a smaller system size $L=128$, and for two different values of the critical force, as indicated. At large forces both curves asymptotically approach the fast-flow limit $V=F$. However, as it is shown in the left-inset, near the depinning transition the velocity strongly depends on the precise value of the critical force.
Therefore, in order to properly analyze the rounding of the depinning transition, averages of small velocities must be carefully defined. To avoid this problem we exploit the access we have to a high-precision value of $F_c$ for each sample~\cite{rosso_depinning_simulation}, as was previously done for the $T=0$ case~\cite{duemmer}. We thus define the reduced velocity and force variables, $v=\overline{\langle \partial_t u(z,t)/F_c \rangle}$ and $f=(F-F_c)/F_c$, respectively. As an example, in the right-inset of Fig.~\ref{f:VdeF} we show the same data than in the left-inset but in the reduced variables.
The improvement of the scaling is obvious. This allows us to define our systematic procedure to extract the physical properties.

For a given disorder realization we first compute the critical-force of the sample and then average the reduced velocity $v$ for a fixed value of the control parameters $f$ and $T$, both set with high precision. Since thermal and disorder averages must be taken in the steady state, we start with a flat initial condition and let the system evolve until the steady state is reached. We then compute $v$ during a sweep over the lateral size $M$. We find that one sweep is also good enough to properly account for the thermal fluctuations of $v$. Disorder average is finally performed by repeating this procedure for $N$ samples of a given system size. In the following $v$ represents the final, thermal and disorder, averaged value of the steady-state velocity.

Since the critical force distribution and the geometry of the critical configuration depends on the relation between $L$ and $M$ for a periodic system~\cite{bolech_critical_force_distribution}, it is important to properly set the aspect-ratio of the system before attempting any finite-size analysis of the data. An efficient choice \cite{bolech_critical_force_distribution} is to work with $M=L^{\zeta_{dep}}$, where $\zeta_{dep}$ is the roughness of the line at $T=0$ depinning. We take the value  $\zeta_{dep}=1.25$ determined numerically \cite{rosso_roughness_at_depinning,duemmer}.
This choice ensures on one hand that $M$ is large enough for periodicity effects to be absent and decorrelation at the largest length-scales to take place in one sweep, and on the other hand that $M$ is small enough to have no more that one dominant configuration~\cite{kolton_depinning_zerot2} controlling the small velocity regime in each sample. Otherwise this would complicate drastically the analysis of thermal effects at low temperatures. In the following, we present results with $L=128,\,256,\,512,\,1024$ and $2048$, with $M=430,\,1024,\,2436,\,5792$ and $13770$, respectively.

\begin{figure}
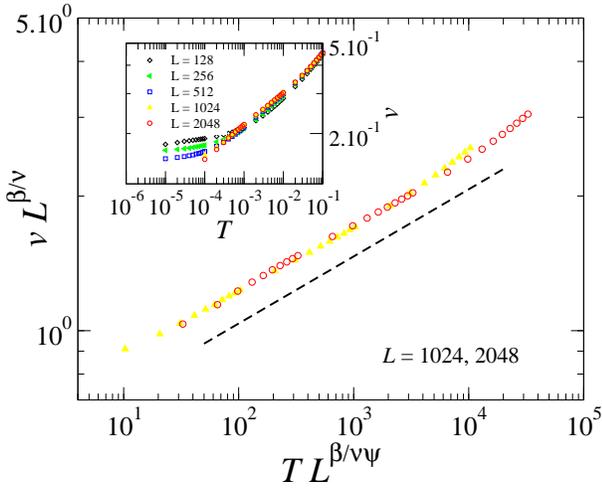

\onefigure[width=8cm]{vdeT-scal.eps} \caption{Finite-size scaling of the
velocity. Main panel: rescaled velocity for $L=1024$ and  $L=2048$ from which
we have extracted the thermal exponent $\psi=0.15 \pm 0.01$ using
eq.~(\ref{eq:vTL}). The power law fit is represented by the continuous dashed
line. The inset shows $v$ for all the  system sizes. The other, previously
known, $T=0$ critical exponents used in the scaling are $\beta = 0.33$ and
$\nu = 1.33$.} \label{f:vdeT-scal}
\end{figure}

The inset of Fig.~\ref{f:vdeT-scal} shows $v$ against $T$ at $F_c$, {\it i.e.} $f=0$, for different system sizes. Strong finite size effects can clearly be seen at the smallest temperatures, where the motion becomes increasingly more correlated. Moreover, we find that finite size effects are appreciable up to $L=512$ in all the temperature range analyzed in Fig.~\ref{f:vdeT-scal}. However, for $L \geq 1024$ all curves collapse very well for $T>5.10^{-4}$, as can be seen by comparing $L=1024$ and $L=2048$. This data is averaged over $N=1000$ samples for $L=128,\,256$, $N=500$ samples for $L=512,\,1024$, and $N=100$ for $L=2048$.
In analogy with critical phenomena, one can assume that the {\it steady-state} velocity, which represents the order parameter of the depinning transition~\cite{fisher_depinning_meanfield}, is an homogeneous function of the ``state variables'', $f$ and $T$, although only for positive $f$~\cite{kolton_depinning_zerot2}. This is consistent with the existence of a growing correlation length controlled by the velocity. With this assumption, scaling arguments lead to universal functions allowing to describe different finite-size regimes. In the zero-temperature case such scaling relation can be written in terms of the system size and the force as
\begin{equation}
\label{eq:vfL}
v \sim L^{-\beta/\nu}\,g\left(f \, L^{1/\nu} \right).
\end{equation}
where the scaling function $g(x)$ is such that $g(x) \sim 1$ for $x \ll 1$ and $g(x) \sim x^\beta$ for $x \gg 1$. This scaling relation accounts for finite-size effects, allowing to obtain a precise estimate~\cite{duemmer} of the depinning exponent $\beta$. It has also been used successfully to analyze the transient critical dynamics of driven elastic lines~\cite{kolton_short_time_exponents}.
Under equivalent assumptions we can write the following scaling relation for the system at the critical force $f=0$ at finite temperature~\cite{narayan_fisher_cdw,middleton_CDW_thermal_exponent},
\begin{equation}
\label{eq:vTL}
v \sim L^{-\beta/\nu}\,h\left(T \, L^{\beta/(\psi \nu)} \right),
\end{equation}
with a scaling function $h(x)$ behaving as $h(x) \sim 1$ for $x \ll 1$ while $h(x) \sim x^\psi$ for $x \gg 1$. We use (\ref{eq:vTL}) to analyze the data presented in the inset of Fig.~\ref{f:vdeT-scal}. The results are presented in the main panel of Fig.~\ref{f:vdeT-scal}. A good collapse is obtained only for the two largest system sizes $L=1024,2048$, while a poor collapse is obtained for system sizes up to $L=512$, due to the large finite-size effects which reduce significantly the region for thermal scaling. Using the values $\beta=0.33$, $\nu=1.33$~\cite{kolton_short_time_exponents} we extract the thermal exponent $\psi$ by fitting the power law $v \sim T^\psi$ for these two largest systems sizes. The best fit is obtained with $\psi = 0.15 \pm 0.01$. The power-law behavior is indicated by the continuous dashed-line in Fig.~\ref{f:vdeT-scal}.

The scaling relation~(\ref{eq:vTL}) strongly suggests the existence of a growing dynamical correlation length $\xi_T$ at depinning, such that $\xi_T \sim T^{-\nu \psi / \beta}$ at the critical force. We test this hypothesis by measuring the temperature-dependent structure factor $S(q)$ of the interface. The structure factor is defined by
\begin{equation}
\label{eq:Sdeq}
S(q) = \overline{ \left\langle \left| \frac{1}{L} \sum_{j=0}^{L-1} u_j(t) \, e^{-iqj} \right| \right\rangle },
\end{equation}
where $q=2\pi n/L$, with $n=1,...,L-1$. From simple dimensional analysis one infers that for small $q$, $S(q) \sim q^{-(1+2\zeta)}$ for a line with a roughness exponent $\zeta$.

FRG calculations show \cite{chauve_creep} that the large-scale motion of an interface at $T=0$ and finite velocity (i.e. $f>0$) can be described by the Edwards-Wilkinson model with an effective temperature determined by the velocity and by the disorder strength. At $T=0$ the corresponding crossover length $\xi_f$, diverges as $f \rightarrow 0+$ (see however Ref.~\cite{kolton_depinning_zerot2} for the situation below threshold) and separates two roughness regimes \cite{chauve_creep,duemmer}: at length scales lower than $\xi_f$, the roughness scales with the depinning exponent $\zeta_{dep}$, while at length scales larger than $\xi_f$ it scales with a purely thermal roughness exponent $\zeta_T=1/2$, with $\zeta_T < \zeta_{dep}$. It is thus natural to check whether the correlation length $\xi_T$ which controls the rounding of the depinning transition at $f=0$ and $T>0$ has the same geometrical interpretation.
In such a case, for $f=0$ and $T>0$, the structure factor would scale as
\begin{equation}
\label{eq:Sq-scal}
S(q) \sim T^{-\nu\psi(1+2\zeta_{dep})/\beta} \; s\left( q \, T^{-\nu\psi/\beta} \right).
\end{equation}
where the scaling function $s(x)$ behaves as $s(x) \sim x^{-(1+2\zeta_T)}$ for $x \ll 1$ and $s(x) \sim x^{-(1+2\zeta_{dep})}$ for $x \gg 1$.
\begin{figure}
\onefigure[width=8cm]{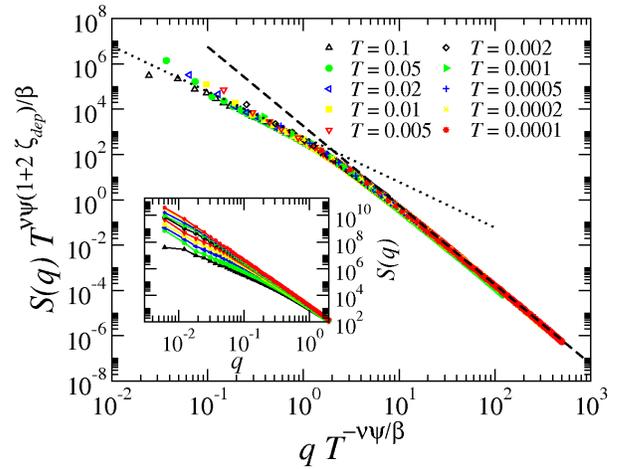}
\caption{Scaling of the finite-temperature structure factor for $L=1024$. The inset shows $S(q)$ for different temperatures, as indicated, and the main panel shows its rescaled form using eq.~(\ref{eq:Sq-scal}). We have used the exponent $\psi$ extracted from the velocity in fig.~\ref{f:vdeT-scal}, and the known roughness exponents $\zeta_T=0.5$ and $\zeta_{dep}=1.25$. Dotted and dashed lines correspond to the asymptotic behaviors $x^{-(1+2\zeta_T)}$ and $x^{-(1+2\zeta_{dep})}$, respectively, with $x=q\,T^{-\nu\psi/\beta}$. The characteristic thermally-induced length-scale $\xi_T \sim T^{-\nu\psi/\beta}$ separates these two roughness regimes.}
\label{f:Sdeq}
\end{figure}

The main panel of Fig.~\ref{f:Sdeq} shows the rescaling of the structure function $S(q)$ according to Eq.~(\ref{eq:Sq-scal}) for different temperatures (the unscaled data is shown in the inset). The system size used is $L=1024$ and the range of temperatures analyzed covers well the power law regime of $v$ from which we have measured the thermal exponent $\psi$. To rescale $S(q)$ we have used all the previously known exponents: $\psi$ was extracted from the velocity in Fig.~\ref{f:vdeT-scal}, $\zeta_T=0.5$ and $\zeta_{dep}=1.25$ from $T=0$ depinning simulations~\cite{rosso_roughness_at_depinning,duemmer}.
As it can be observed, the collapse of curves is excellent in the three temperature decades analyzed.
Fig.~\ref{f:vdeT-scal} and Fig.~\ref{f:Sdeq} thus give strong support to the view that the scaling behavior at the depinning transition, both for $f>0$, $T=0$ and the $f=0$, $T>0$ cases, is controlled by the {\it same} velocity-dependent correlation length, having identical geometrical interpretations. This length diverges with the velocity as $\xi \sim v^{-\nu/\beta}$ regardless of how the $v=0$ critical point is approached ($f\rightarrow 0+$ at $T=0$ or $T\rightarrow 0$ at $f=0$).

Let us now compare our results with previous analysis of the thermal rounding.
In ref.~\cite{chen_marchetti}, the authors gave a first estimate for the thermal exponent $\psi$ of an elastic string in a random medium. Instead of dealing with a finite-size system, they simulated the driven interface in a semi-infinite medium with $M \to \infty$ by dynamically generating disorder in a small region of the sample around the moving system. By fitting the resulting time-averaged velocity-force characteristics they have calculated the critical force, and the critical exponents. The value they estimated for the thermal exponent, $\psi=0.16$~\cite{chen_marchetti}, is very close to our present estimate, $\psi =0.15$. However, their numerical procedure also gives $\beta \approx  0.24$, which is much lower than the one obtained in more recent numerical simulations, $\beta \approx 0.33$ \cite{duemmer,kolton_short_time_exponents} and analytical calculations \cite{chauve_creep}, prompting for questions on the reliability of such a semi-infinite procedure to obtain the critical exponents.
The thermal exponent $\psi$ has been also measured in numerical simulations of domain wall motion in the random field Ising model, by fitting various parameters of the $V(F,T)$ curves in order to obtain universal functions. The reported value for this indirect measure is $\psi \approx 0.2$ in (1+1) dimensions~\cite{nowak_thermal_rounding}, higher than our value. Although our study rests on random-bond disorder, it has been shown \cite{chauve_creep,rosso_correlator_RB_RF_depinning} that for the $T=0$ dynamics random-bond disorder and random-field are in the same universality class, contrarily to the statics. One would thus naively expect the same thermal exponent as the one found in Ref.~\cite{nowak_thermal_rounding}, assuming that anharmonic corrections to the elasticity, present in the random fied Ising model but absent in our model, can not change the value of $\psi$. If that is the case for the random field Ising model, the difference in the thermal exponent can be attributed to the numerical limitations in \cite{nowak_thermal_rounding} compared to our method, where the control parameter $F-F_c$ can be determined with high accuracy, for obtaining this exponent.

More recently, a finite temperature study of the depinning transition in a model of extremal activated dynamics~\cite{vandembroucq_thermal_rounding_extremal_model} was reported. Analyzing the geometry of the line at depinning with this artificial dynamics gives a characteristic length $l \sim T^{-0.95}$ which separates the $\zeta_{dep}=1.25$ and $\zeta_T=0.5$ regimes of roughness and which can be also associated with the distribution of subcritical forces along the front. If $l$ was the velocity-dependent dynamical correlation length of the depinning transition $\xi$, this result would imply $v\sim T^{0.24}$, yielding a value of $\psi=0.24$, which is much higher than ours and previous reported values. So most likely this artificial dynamics does not allow to infer the velocity for the depinning problem.

One of the interests in trying to obtain an accurate determination of the
rounding exponent is to determine which scaling law governs it and whether it
is an independent exponent from the $T=0$ depinning exponents or not. Indeed,
in the simpler case of a particle in a one dimensional potential, the value of
the thermal exponent is related to the first-passage-time problem of
overcoming, by thermal fluctuations, the vanishing barrier at the depinning or
saddle-node
bifurcation~\cite{Bishop_josephson_junction,Colet_marginal_thermal_escape}.
This leads to the relation $\psi=\beta/(2-\beta)=1/3$ with $\beta=1/2$ for any
analytical potential in one dimension with non-vanishing second derivative at
the critical point. For the case of the interface no solid determination of
the rounding exponent exists. One
proposal~\cite{tang_stepanow_thermal_rounding} for the rounding exponent is
$\psi = \beta/(1+2\beta)$. This relation, with $\beta=1/3$ would lead to
$\psi=1/5$ which would be much higher than our numerical estimate, although it
compares well with the random field Ising model value. It would also work
poorly for the $d=2$ case of the random field Ising model (using $\beta=2/3$
it would give $\psi = 2/7$ instead of the
measured~\cite{roters_thermal_rounding1} $\psi= 0.42$) and seems thus to be
ruled out. For charge-density-waves it was proposed~\cite{narayan_fisher_cdw}
using mean-field theory that $\psi = 3\beta /2$. This law agreed with
numerical simulations for finite-dimension charge-density-waves
models~\cite{middleton_CDW_thermal_exponent}. Using the analogy with standard
critical phenomena (viewing $F-F_c$ as $T-T_c$, $V$ as a magnetization, and
$T$ as an external magnetic field), a scaling relation for the thermal
exponent can be obtained by using the standard hyperscaling
relation~\cite{roters_thermal_rounding1}, leading to
$\psi=\beta/[(d+1)\nu-\beta]$, where $d$ is the internal dimension of the
interface ($d=1$ in the present case). Such an estimate gives good results for
the $d=2$ rounding exponents for the random field Ising
model~\cite{roters_thermal_rounding1}: using $\beta=2/3$ and $\nu=3/4$ it
predicts $\psi=8/19\approx 0.421$ which compares well with the numerical value
$\psi= 0.42$. For our one-dimensional case, it would predicts the value
$\psi=1/7 \approx 0.143$, lower but very close to our numerical value. This
relation seems thus empirically quite good. Of course, this approach is purely
phenomenological as there is not any clear justification for using the
equilibrium scaling relations for non-equilibrium dynamical transitions.

A rigorous derivation is thus clearly needed. Although in principle the finite
temperature FRG calculation~\cite{chauve_creep} allows to reach the thermal
exponent, the equations are quite complicated and the involved analytical
approach has not been accomplished so far. In that respect our results provide
an important clue by showing that the scaling properties near the critical
point $f=T=0$ are controlled by the same velocity-dependent correlation length
as for the $T=0$, $f>0$ case. They show that the large-scale geometry is also
the same in both cases, regardless the origin of the steady-state velocity. In
the FRG this implies that the flow of the friction/velocity (the parameter
$\lambda$ in \cite{chauve_creep}) is identical to the zero-temperature case,
providing an independent confirmation of the hypothesis used to study the
$T=0$ flow, that the rounding of the cusp occurs at a scale $\rho_\lambda$
which is negligible compared to $\lambda$ close to the depinning
\cite{chauve_creep}. Note that since $v \sim T^\psi$ and $\psi < 1$ the
velocity is indeed large compared to the temperature. This makes it likely
that the whole rounding of the cusp is still controlled by the velocity,
although in principle it could be also possible that, even if the temperature
is smaller than $\lambda$, it is larger that the velocity cusp-rounding scale,
$\rho_\lambda$. Our result thus urge for a reexamination and a further
analysis, either analytical or numerical, to check these possible scenarios
and to extract the rounding exponent directly from the FRG
calculation~\cite{bustingorry_thermal_rouding_long}. Finally, let us point out
that, although the results here presented correspond strictly to the
steady-state evolution of the driven interface, we have also found that the
value of $\psi$ agrees well with the ones obtained from the analysis of the
non-steady short-time relaxation of the interface at $f=0$ and $T>0$
~\cite{bustingorry_thermal_rouding_long}.

Our results are directly relevant for recent experiments on the domain wall motion in ferromagnetic thin films~\cite{metaxas_depinning_thermal_rounding}. Indeed, in these experiments both the thermal exponent $\psi$ and the geometrical properties such as the dynamical correlation length $\xi$ and the roughness exponents $\zeta_T$ and $\zeta_{dep}$ could be obtained by imaging the structure of the moving domain walls. This would allow to check the current picture of the depinning transition as a collective phenomenon. It is also worth stressing that our study, and in particular our finite-size effects analysis, is relevant for other model systems with system-size limitations, such as the recently reported simulations on the temperature dependence of the flux lines dynamics in high-temperature superconductors~\cite{luo_thermal_rounding_flux_lines}.

In conclusion we have studied the thermal rounding of the depinning transition by analyzing both the steady-state velocity and geometry of a one-dimensional interface moving in a two dimensional random medium. We have obtained the thermal exponent $\psi=0.15 \pm 0.01$ by directly extracting the temperature dependence of the velocity at the critical force, $v\sim T^{\psi}$. As it was recently done in Ref~\cite{duemmer} for the $T=0$ case, we have exploited the easy access to the exact critical force for {\it each} disorder realization by using the powerful Rosso-Krauth algorithm~\cite{rosso_depinning_simulation,rosso_roughness_at_depinning}. This allowed us to eliminate the statistical uncertainty in the control parameter induced by the sample to sample fluctuations of the critical force, which is present in all previous Langevin-dynamics numerical approaches to the thermal-rounding problem. Moreover, we have shown explicitly that the value of $\psi$ is consistent with the existence of a velocity-dependent correlation length $\xi$ separating two regimes of roughness, and we find that $\xi$ diverges as $v\rightarrow 0$ the same way, regardless we approach the depinning threshold $f=T=0$ from positive forces, or from the temperature axis. Our results are relevant for recent experiments on domain wall motion in ferromagnetic thin films~\cite{metaxas_depinning_thermal_rounding}.

\acknowledgments
We thank A. Rosso for illuminating discussions. This work was supported in part by the Swiss National Science Foundation under
MaNEP and Division II.

\bibliography{tfinita,totphys}
\bibliographystyle{iopart-num}
\end{document}